\documentclass[twocolumn]{svjour3}
\usepackage{graphicx}
\usepackage{amsmath,amssymb}

\journalname{Eur. Phys. J. C}

\begin{document}

\title{Investigation of hidden-charm pentaquarks with strangeness $S=-1$}

\author{Xiaohuang Hu \and Jialun Ping\thanks{Corresponding author: jlping@njnu.edu.cn}}

\institute{Department of Physics and Jiangsu Key Laboratory for Numerical
Simulation of Large Scale Complex Systems, Nanjing Normal University, Nanjing 210023, P. R. China}

\titlerunning{Investigation of hidden-charm pentaquarks in single strangeness sector}
\authorrunning{Hu,Ping}

\maketitle

\begin{abstract}
Recently, a new hidden-charm pentaquark state $P_{cs}(4459)$ was reported by the LHCb Collaboration. Stimulated by the fact that all hidden-charm pentaquark states in $S=0$ systems were successfully studied by the chiral quark model, we extended this study to the $S=-1$ systems. All possible states with quantum numbers $IJ^P=0(\frac{1}{2})^-$, $0(\frac{3}{2})^-$, $0(\frac{5}{2})^-$, $1(\frac{1}{2})^-$, $1(\frac{3}{2})^-$ and $1(\frac{5}{2})^-$ have been investigated. The calculation results shows that the newly observed state $P_{cs}(4459)$ can be explained as $\Xi_c \bar{D}^*$  molecular state and the quantum numbers are $0(\frac{1}{2})^-$. In addition, we also find other molecular states $\Xi_c \bar{D}$, $\Xi_c^* \bar{D}$ and $\Xi_c' \bar{D}^*$. It is worth mentioning that $\Xi_c \bar{D}^*$ can form a two-peak structure from states in system $0(\frac{1}{2})^-$ and $0(\frac{3}{2})^-$. The decay width of all molecular states is given with the help of real scaling method. These hidden-charm pentaquark states is expected to be further verified in future experiments.
\end{abstract}

\section{Introduction}
Since $\Theta^+$ was proposed in 2003~\cite{1,2}, the search of pentaquark states has always been a hot topic. A large number of exotic hadronic states,
``$XYZ$" states which associated with tetraquark, have been discovered in experiments over the past decade. Naturally the observation of pentaquark
state is expected. The important and prominent event arrived in 2015, the $P_c^+$ states $P_c^+(4380)$ and $P_c^+(4450)$, which are the strongest candidates
for pentaquark states were observed by the LHCb Collaboration in the $J/\psi p$ invariant mass spectrum of decay channel
$\Lambda_b\rightarrow J\psi pK$~\cite{3}. And then in 2019, the LHCb Collaboration reanalysed the same process with more data and updated their results,
which shows that $P_c^+(4450)$ can split into two structures, $P_c^+(4440)$ and $P_c^+(4457)$, while $P_c^+(4312)$ is identified as a new structure~\cite{4}.
These $P_c^+$ states all have a common feature that they are very close to baryon-meson thresholds, which led to one of the most popular interpretation of
these $P_c^+$ states as molecular states~\cite{5,6,7,8,9,10,11}. Other possible explanations also exist, such as the compact pentaquark states~\cite{12,13}.
In addition, the decay properties and width of these $P_c^+$ states also have been investigated in Ref.~\cite{14}.

Very recently, the LHCb Collaboration reported a new pentaquark state with strangeness, $P_{cs}^+(4459)$ in the $J/\psi \Lambda$ invariant mass spectrum
of decay channel $\Xi_b\rightarrow J\psi K^-\Lambda$~\cite{15}. The mass and decay width of $P_{cs}^+(4459)$ are,
\begin{eqnarray}
&& M = 4458.8 \pm 2.9 ^{+4.7}_{-1.1} ~\mbox{MeV}, \nonumber \\
&& \Gamma = 17.3 \pm 6.0^{+8.0}_{-5.7} ~\mbox{MeV},  \nonumber  \nonumber
\end{eqnarray}
while its spin and parity are unknown. Like other $P_c^+$ pentaquark states, the mass of $P_{cs}^+(4459)$ is very close to threshold of
$\Xi_c \bar{D}^*$. Thus, some phenomenological models have been used to investigate whether $P_{cs}^+(4459)$ can be explained as a $\Xi_c \bar{D}^*$
molecular state, such as QCD sum rule~\cite{16,17}, coupled channel unitary approach~\cite{18}, effective field theory~\cite{19} and so
on~\cite{20,21,22,23}. Other structures of $P_{cs}$ are also proposed, for example, diquark-diquark-antiquark structure \cite{Azizi}.
Actually, there are many theoretical work on $P_{cs}$ state before the experiment~\cite{24,25,26,27}.

It is important to highlight here that, before the LHCb's discovered several $P_c^+$ pentaquark states, their existence was predicted \cite{JJWu1,JJWu2,ZCYang},
and our quark model calculation also gave a good description and prediction~\cite{28}. $P_{c}^+(4312)$, $P_{c}^+(4440)$ and $P_{c}^+(4459)$ were described as baryon-meson molecular states with the
$IJ^P=\frac{1}{2}(\frac{1}{2})^- ~\Sigma_c \bar{D}$, $\frac{1}{2}(\frac{1}{2})^-~\Sigma_c \bar{D}^*$ and $\frac{1}{2}(\frac{3}{2})^-~\Sigma_c \bar{D}^*$,
respectively. A lot of subsequent theoretical work also supports our conclusion~\cite{9,10,11,29}. Thus, it is natural to extend herein such study
from $P_{c}$ states to the $P_{cs}$ states. Since the quark model was proposed  by M. Gell-Mann and G. Zweig in 1964 respectively~\cite{30,31},
it has become the most common approach to study the multiquark system as it evolves. In this work, the constituent chiral quark model (ChQM) will still
be employed to investigate $qqsc\bar{c}$ systems ($q$ stands for $u$ or $d$) corresponding to $P_{cs}$ states. To calculate accurately the results
of each possible system, Gaussian expansion method (GEM)~\cite{32}, an accurate and universal few-body calculation method is adopted. The GEM is very
suitable for the calculation of few-body systems. Within this method, the orbital wave functions of all relative motions of the systems are expanded
by gaussians. After considering all possible configuration of color, spin and flavor degrees of freedom, we can identify
the structures of the system. Finally, with the help of ``real scaling method", we can confirm the genuine five-quark resonances and their
respective decay widths along.

The paper is organized as follow. After introduction, details of ChQM and GEM are introduced in Section II.
In Section III, we present the method of finding and calculating the decay width of the genuine resonance state (``real scaling method"),
and then we show the results with analysis and discussion of $P_{cs}$ structure. Finally, We give a brief summary of this work in the last section.

\section{CHIRAL QUARK model and wave functions}

In this paper, ChQM is employed to investigate the $P_{cs}$ states. The model has become one of the most common approaches to describe
hadron spectra, hadron-hadron interactions and multiquark states~\cite{33}. In this model, in addition to one-gluon exchange (OGE), the massive
constituent quarks also interact with each other through Goldstone boson exchange. Besides, the color confinement and the scalar $\sigma$ meson
(chiral partner, acting on $u$ and $d$ quark only) exchange are also introduced. More details of this model can be found in Ref.~\cite{33,34}.
The Hamiltonian of ChQM is given as follows:
\begin{eqnarray}
H & = & \sum_{i=1}^{n}\left( m_i+\frac{p^2_i}{2m_i}\right)-T_{CM}
      + \sum_{j>i=1}^{n} V_{ij} \\
V_{ij} & = & V^{C}_{ij}+V^{G}_{ij}+V^{\chi}_{ij}+V^\sigma_{ij} , \\
V^{C}_{ij} & = & \boldsymbol{\lambda}_i^c\cdot \boldsymbol{\lambda}_j^c
   \left[-a_c (1-e^{-\mu_cr_{ij}})+\Delta \right] ,
\end{eqnarray}
\begin{eqnarray}
V^{G}_{ij} & = & \frac{\alpha_s}{4} \boldsymbol{\lambda}_i^c \cdot \boldsymbol{\lambda}_j^c
   \left[ \frac{1}{r_{ij}}-\frac{1}{6m_im_j} \boldsymbol{\sigma}_i\cdot \boldsymbol{\sigma}_j
   \frac{e^{-r_{ij}/r_0(\mu)}}{r_{ij}r^2_0(\mu)}\right] , \\
   & & r_0(\mu)=\hat{r}_0/\mu,~~\alpha_{s} =
   \frac{\alpha_{0}}{\ln(\frac{\mu^2+\mu_{0}^2}{\Lambda_{0}^2})}.  \\
V^{\chi}_{ij} & = & v_{\pi}({{\bf r}_{ij}})\sum_{a=1}^{3}
	\lambda_i^a \lambda_j^a+v_{K}({{\bf r}_{ij}})\sum_{a=4}^{7}
	\lambda_i^a \lambda_j^a) \nonumber \\
& & +v_{\eta}({{\bf r}_{ij}})
    [\cos\theta_{P}(\lambda_i^8 \lambda_j^8)-\sin\theta_{P}] ,  \\
v^{\chi}_{ij} & = & \frac{g^2_{ch}}{4\pi}\frac{m^2_\chi}{12m_im_j}
	\frac{\Lambda^2_\chi}{\Lambda^2_\chi-m^2_\chi}m_\chi  \nonumber \\
& &	\left[ Y(m_{\chi}r_{ij})-\frac{\Lambda^3_\chi}{m^3_\chi}Y (\Lambda_{\chi}r_{ij}) \right]
	(\boldsymbol{\sigma}_i \cdot \boldsymbol{\sigma}_j),  \\
& & \chi=\pi,K,\eta , \\
V^{\sigma}_{ij} & = & -\frac{g^2_{ch}}{4\pi} \frac{\Lambda^2_\sigma m_\sigma}{\Lambda^2_\sigma-m^2_\sigma}
	\left[ Y(m_{\sigma}r_{ij})-\frac{\Lambda_\sigma}{m_\sigma}Y(\Lambda_{\sigma}r_{ij})\right].
\end{eqnarray}
where $T_{cm}$ is the kinetic energy of the center-of mass motion and $\mu$ is the reduced mass between two interacting quarks. Only the central parts
of the interactions are given here because we are interested in the low-lying states of the multiquark system. $\boldsymbol{\sigma}$
represents the SU(2) Pauli matrices; $\boldsymbol{\lambda}^c$ and $\boldsymbol{\lambda}$ represent the SU(3) color and flavor Gell-Mann matrices
respectively; $\alpha_s$ denotes the strong coupling constant of one-gluon exchange and $Y(x)$ is the standard Yukawa functions.
Because it is difficult to use the same set of parameters to have a good description of baryon and meson spectra simultaneously, we treat the strong
coupling constant of one-gluon exchange with different values for quark-quark and quark-antiquark interacting pairs.

The model parameters are listed in Table I, and the calculated baryon and meson masses are presented in the Table II with the experimental values.
From the calculation,most of the results are close to experimental values except for the $\Lambda_c$ and $J/\psi$. In the follow-up calculation, we find that the molecular state corresponding to these two hadrons are open channels and these calculation errors do not cause mass inversion, so these errors do not affect our final results.
\begin{table}[h]
\caption{Quark model parameters}
\begin{tabular}{c|cc cc} \hline
Quark masses       &$m_u$=$m_d\;(MeV)$   &~~~~313\\
                   &$m_s\;(MeV)$  &~~~~555\\
                    &$m_c\;(MeV)$  &~~~~1780\\   \hline
                   &$\Lambda_\pi=\Lambda_\sigma~$ (fm$^{-1}$)  &~~~~4.20\\
                   &$\Lambda_\eta=\Lambda_K~$ (fm$^{-1}$)      &~~~~5.20\\
                   &$m_\pi$ (fm$^{-1}$)  &~~~~0.70\\
Goldstone bosons   &$m_K$ (fm$^{-1}$)  &~~~~2.51\\
                   &$m_\eta$ (fm$^{-1}$)  &~~~~~2.77\\
                   &$m_\sigma$ (fm$^{-1}$)  &~~~~~3.42\\
                   &$g^2_{ch}/(4\pi)$  &~~~~0.54\\
                   &$\theta_P(^\circ)$  &~~~~-15\\  \hline
                   &$a_c$ (MeV)  &~~~~280.3\\
     Confinement       &$\mu_c$ (fm$^{-1})$  &~~~~0.863\\
                    &$\Delta$ (MeV)  &~~~~115.0\\   \hline

                 &$\hat{r}_0~$(MeV~fm)  &~~~~49.3\\
                    &$\alpha_{uu}$  &~~~~0.623/0.924\\
        OGE           &$\alpha_{us}$  &~~~~0.915/-\\
                    &$\alpha_{uc}$  &~~~~0.900/0.765\\
                   &$\alpha_{sc}$  &~~~~0.710/0.633\\
                   &$\alpha_{cc}$  &~~~~-/0.5\\ \hline
\end{tabular}
\end{table}

\begin{table}[h]
\caption{The masses of ground-state baryons and mesons(unit: Mev)}
\begin{tabular}{ccccccc}
\hline
 &$\Lambda$~ &$\Sigma$~ &$\Sigma^*$~ &$\Lambda_c$~ &$\Sigma_c'$~  &$\Sigma_c^*$~   \\ \hline
CHQM~ &1114~ &1243~ &1404~ &2184~ &2453~ &2529~  \\
Expt~ &1116~ &1189~ &1385~ &2286~ &2455~ &2520~  \\ \hline
 &$\Xi_c$~  &$\Xi_c'$~ &$\Xi_c^*$~ \\ \hline
CHQM~  &2460~ &2580~ &2653~  \\
Expt~  &2471~ &2589~ &2645~  \\ \hline

&$\pi$~ &$\rho$~ &$D$~ &$D^*$~ &$D_s$~ &$D_s^*$  \\ \hline
CHQM~ &140~ &698~ &1858~ &2023~ &1964~ &2156~ \\
Expt~ &140~ &775~ &1864~ &2007~ &1968~ &2112~ \\
 &$\eta_c$~ &$J/\psi$  \\ \hline
CHQM~  &2984~ &3182~ \\
Expt~  &2984~ &3097~ \\
\hline
\end{tabular}
\end{table}

In the following, the wave functions for the five-quark systems are constructed and the eigen-energy is obtained by solving the Schr\"{o}dinger equation.
The wave function of the system consists of four parts: orbital, spin, flavor and color. The wave function of each part is constructed in two steps,
first construct the wave function of three-quark cluster and quark-antiquark cluster, respectively, then coupling two clusters wave functions to form the
complete five-body one.
In the following, the wave functions for $(qqs)(\bar{c}c)$ configuration is written down, the wave functions for other configuration can be obtained
by exchange the indices of particles. The indices of particles $q,q,s,\bar{c},c$ are 1,2,3,4,5. The wave functions for $(qqc)(\bar{c}s)$ are obtained
by exchange the particle indices $3\leftrightarrow 5$ as an example.

The first part is orbital wave function. A five-body system have four relative motions so it is written as follows.
\begin{equation}
\psi^x_{LM_L}=\left[ \left[ \left[
  \psi_{n_1l_1}(\mbox{\boldmath $\rho$})\psi_{n_2l_2}(\mbox{\boldmath $\lambda$})\right]_{l}
  \psi_{n_3l_3}(\mbox{\boldmath $r$}) \right]_{l^{\prime}}
  \psi_{n_4l_4}(\mbox{\boldmath $R$}) \right]_{LM_L},
\end{equation}
where the Jacobi coordinates are defined as follows,
\begin{eqnarray}
{\mbox{\boldmath $\rho$}} & = & {\mbox{\boldmath $x$}}_1-{\mbox{\boldmath $x$}}_2, \nonumber \\
{\mbox{\boldmath $\lambda$}} & = & (\frac{{m_1\mbox{\boldmath $x$}}_1+{m_2\mbox{\boldmath $x$}}_2}{m_1+m_2})-{\mbox{\boldmath $x$}}_3,  \nonumber \\
{\mbox{\boldmath $r$}} & = & {\mbox{\boldmath $x$}}_4-{\mbox{\boldmath $x$}}_5,  \\
{\mbox{\boldmath $R$}} & = & (\frac{{m_1\mbox{\boldmath $x$}}_1+{m_2\mbox{\boldmath $x$}}_2
  +{m_3\mbox{\boldmath $x$}}_3}{m_1+m_2+m_3})
  -(\frac{{m_4\mbox{\boldmath $x$}}_4+{m_5\mbox{\boldmath $x$}}_5}{m_4+m_5}). \nonumber
\end{eqnarray}
$\boldsymbol{x}_i$ is the position of the $i$-th particle. Then we use a set of gaussians to expand the radial part
of the orbital wave function which is shown below,
\begin{eqnarray}
\psi_{lm}(\mathbf{r})=\sum^{n_{max}}_{n=1}c_{nl}\phi^{G}_{nlm}(\mathbf{r})
\end{eqnarray}
\begin{eqnarray}
\phi^{G}_{nlm}(\mathbf{r})=\emph{N}_{nl}r^{l}e^{-\nu_{n}r^{2}}\emph{Y}_{lm}(\hat{\mathbf{r}})
\end{eqnarray}
where $N_{nl}$ is the normalization constant,
\begin{eqnarray}
\emph{N}_{nl}=\left(\frac{2^{l+2}(2\nu_{n})^{l+3/2}}{\sqrt\pi(2l+1)!!}\right)^{\frac{1}{2}},
\end{eqnarray}
and $c_{nl}$ is the variational parameter, which is determined by the dynamics of the system. The Gaussian size
parameters are chosen according to the following geometric progression:
\begin{eqnarray}
\nu_{n}=\frac{1}{r^{2}_{n}}, r_{n}=r_{min}a^{n-1}, a=\left(\frac{r_{max}}{r_{min}}\right)^{\frac{1}{n_{max}-1}},
\end{eqnarray}
where $n_{max}$ is the number of Gaussian functions, and $n_{max}$ is determined by the convergence of the results.
In the present calculation, $n_{max}=8$.

The details of constructing flavor, color and spin wave functions of 5-quark system can be found in Ref.~\cite{35}, only the last expressions
are shown here.

Flavor wave functions:
\begin{eqnarray}
 |\chi^{f1}_{0,0} \rangle & = & \frac{1}{\sqrt{2}}(uds\bar{c}c-dus\bar{c}c) \nonumber \\
 |\chi^{f2}_{0,0} \rangle & = &  uus\bar{c} c
\end{eqnarray}

Color wave functions:
\begin{eqnarray}
 |\chi^{c1} \rangle  & = & \frac{1}{\sqrt{18}}(rgb-rbg+gbr-grb+brg-bgr) \nonumber \\
 & & (\bar r r+\bar gg+\bar bb) \nonumber \\
 |\chi^{c2} \rangle  & = &  \frac{1}{\sqrt{192}}\left[ 2(2rrg-rgr-grr)\bar r b \right. \nonumber \\
& + & 2(rgg+grg-2ggr)\bar g b \nonumber \\
& - & 2(2rrb-rbr-brr)\bar r g-2(rbb+brb-2bbr)\bar b g \nonumber \\
& + & 2(2ggb-gbg-bgg)\bar g r+2(gbb+bgb-2bbg)\bar b r \nonumber \\
& + & (rbg-gbr+brg-bgr)(2\bar b b-\bar r r-\bar g g)  \nonumber \\
& + & (2rgb-rbg+2grb-gbr-brg-bgr)(\bar r r-\bar g g)]    \nonumber  \\
 |\chi^{c3} \rangle  & = &  \frac{1}{24}[6(rgr-grr)\bar r b+6(rgg-grg)\bar g b  \\
& - & 6(rbr-brr)\bar r g-6(rbb-brb)\bar b g \nonumber \\
& + & 6(gbg-bgg)\bar g r+6(gbb-bgb)\bar b r \nonumber \\
& + & 3(rbg+gbr-brg-bgr)(\bar r r-\bar g g)  \nonumber \\
& + & (2rgb+rbg-2grb-gbr-brg+bgr)  \nonumber \\
&   & (2\bar b b-\bar r r-\bar g g)]  \nonumber
\end{eqnarray}

Spin wave functions:
\begin{eqnarray}
 |\chi^{\sigma1}_{\frac12,\frac12} \rangle & = & \frac{1}{\sqrt{12}}(2\alpha\alpha\beta\alpha\beta-2\alpha\alpha\beta\beta\alpha
   +\alpha\beta\alpha\beta\alpha \nonumber \\
  & - & \alpha\beta\alpha\alpha\beta+\beta\alpha\alpha\beta\alpha-\beta\alpha\alpha\alpha\beta) \nonumber \\
 |\chi^{\sigma2}_{\frac12,\frac12} \rangle & = & \frac{1}{2}(\alpha\beta\alpha\alpha\beta-\alpha\beta\alpha\beta\alpha+\beta\alpha\alpha\beta\alpha-\beta\alpha\alpha\alpha\beta) \nonumber \\
 |\chi^{\sigma3}_{\frac12,\frac12} \rangle & = & \frac{1}{6}(2\alpha\alpha\beta\alpha\beta+2\alpha\alpha\beta\beta\alpha-\alpha\beta\alpha\alpha\beta-\alpha\beta\alpha\beta\alpha \nonumber \\
& - & \beta\alpha\alpha\beta\alpha-\beta\alpha\alpha\alpha\beta-2\alpha\beta\beta\alpha\alpha-2\beta\alpha\beta\alpha\alpha \nonumber \\
& + & 4\beta\beta\alpha\alpha\alpha) \nonumber \\
 |\chi^{\sigma4}_{\frac12,\frac12} \rangle & = & \frac{1}{\sqrt{12}}(\alpha\beta\alpha\alpha\beta+\alpha\beta\alpha\beta\alpha-\beta\alpha\alpha\alpha\beta-\beta\alpha\alpha\beta\alpha \nonumber \\
& + & 2\beta\alpha\beta\alpha\alpha-2\alpha\beta\beta\alpha\alpha)  \\
 |\chi^{\sigma5}_{\frac12,\frac12} \rangle & = & \frac{1}{\sqrt{18}}(3\alpha\alpha\alpha\beta\beta-\alpha\alpha\beta\alpha\beta
  -\alpha\alpha\beta\beta\alpha \nonumber \\
 & - & \alpha\beta\alpha\alpha\beta-\alpha\beta\alpha\beta\alpha-\beta\alpha\alpha\alpha\beta-\beta\alpha\alpha\beta\alpha  \nonumber \\
& + & \beta\beta\alpha\alpha\alpha+\beta\alpha\beta\alpha\alpha+\alpha\beta\beta\alpha\alpha)  \nonumber \\
 |\chi^{\sigma 1}_{\frac32,\frac32} \rangle & = & \frac{1}{\sqrt{6}}(2\alpha\alpha\beta\alpha\alpha-\alpha\beta\alpha\alpha\alpha
 -\beta\alpha\alpha\alpha\alpha) \nonumber \\
 |\chi^{\sigma 2}_{\frac32,\frac32} \rangle & = & \frac{1}{\sqrt{2}}(\alpha\beta\alpha\alpha\alpha-\beta\alpha\alpha\alpha\alpha) \nonumber
\end{eqnarray}
\begin{eqnarray}
 |\chi^{\sigma 3}_{\frac32,\frac32} \rangle & = & \frac{1}{\sqrt{2}}(\alpha\alpha\alpha\alpha\beta-\alpha\alpha\alpha\beta\alpha) \nonumber \\
 |\chi^{\sigma 4}_{\frac32,\frac32} \rangle & = & \frac{1}{\sqrt{30}}(2\alpha\alpha\beta\alpha\alpha+2\alpha\beta\alpha\alpha\alpha+2\beta\alpha\alpha\alpha\alpha \nonumber \\
& - & 3\alpha\alpha\alpha\alpha\beta-3\alpha\alpha\alpha\beta\alpha)  \nonumber\\
 |\chi^{\sigma 1}_{\frac52,\frac52} \rangle & = & \alpha\alpha\alpha\alpha\alpha , \nonumber
\end{eqnarray}
where $\chi^{c1} $ represents the color wave function of a color singlet-singlet structure, $\chi^{c2}$ and $\chi^{c3} $ represent the color octet-octet
wave functions respectively. The subscripts of $\chi^{f}_{I,I_z}$ ($\chi^{\sigma}_{S,S_z}$) are total isospin (spin) and its third projection.

Finally, the total wave function of the five-quark system is written as:
\begin{equation}
\Psi_{JM_J}^{i,j,k}={\cal A} \left[ \left[  \psi_{L}\chi^{\sigma_i}_{S}\right]_{JM_J} \chi^{fi}_j \chi^{ci}_k \right],
\end{equation}
where the $\cal{A}$ is the antisymmetry operator of the system which guarantees the antisymmetry of the total wave functions when identical
particles exchange. Under our numbering scheme, the antisymmetry operator is
\begin{equation}
 \mathcal{A}  = 1-(12)
\end{equation}

At last, we solve the following Schr\"{o}dinger equation to obtain eigen-energies of the system,
\begin{equation}
H\Psi_{JM_J}=E\Psi_{JM_J},
\end{equation}
with the help of the Rayleigh-Ritz variational principle. The matrix elements of Hamiltonian can be easily obtained if all the orbital angular momenta
are zero, which is reasonable for only considering the low-lying states of five-quark system. It is worthwhile to mention that if the orbital angular
momenta of the system are not zero, it is necessary to use the infinitesimally shifted Gaussian method to calculate the matrix elements~\cite{32}.

\section{Results and discussions}
In this section, we present the calculation results of all low-lying states of the $(uuc)(s \bar c)$, $(uus)(c \bar c)$
and $(usc)(u \bar c)$ five-quark system with all possible quantum numbers $IJ^P=0(\frac{1}{2})^-$, $0(\frac{3}{2})^-$, $0(\frac{5}{2})^-$,
$1(\frac{1}{2})^-$, $1(\frac{3}{2})^-$ and $1(\frac{5}{2})^-$ in ChQM. All the orbital angular momentum of the system is treated as zero and
the corresponding parity is negative. In addition, a stability method to identify genuine resonance states, real-scaling method~\cite{35,36,37},
is employed. In this method, the Gaussian size parameter $r_{n}$ for the basis functions between baryon and meson clusters for the color-singlet
channels is scaled by multiplying a factor $\alpha$: $r_{n} \longrightarrow \alpha r_{n}$. As a result, a genuine resonance will act as an
avoid-crossing structure (see Fig.~1) with the increasing of $\alpha$, while other continuum states will fall off towards its threshold.
If the avoid-crossing structure is repeated periodically as $\alpha$ increase, then the avoid-crossing structure is a genuine resonance~\cite{38}.
\begin{figure}[h]
  \centering
  \includegraphics[width=9cm,height=7cm]{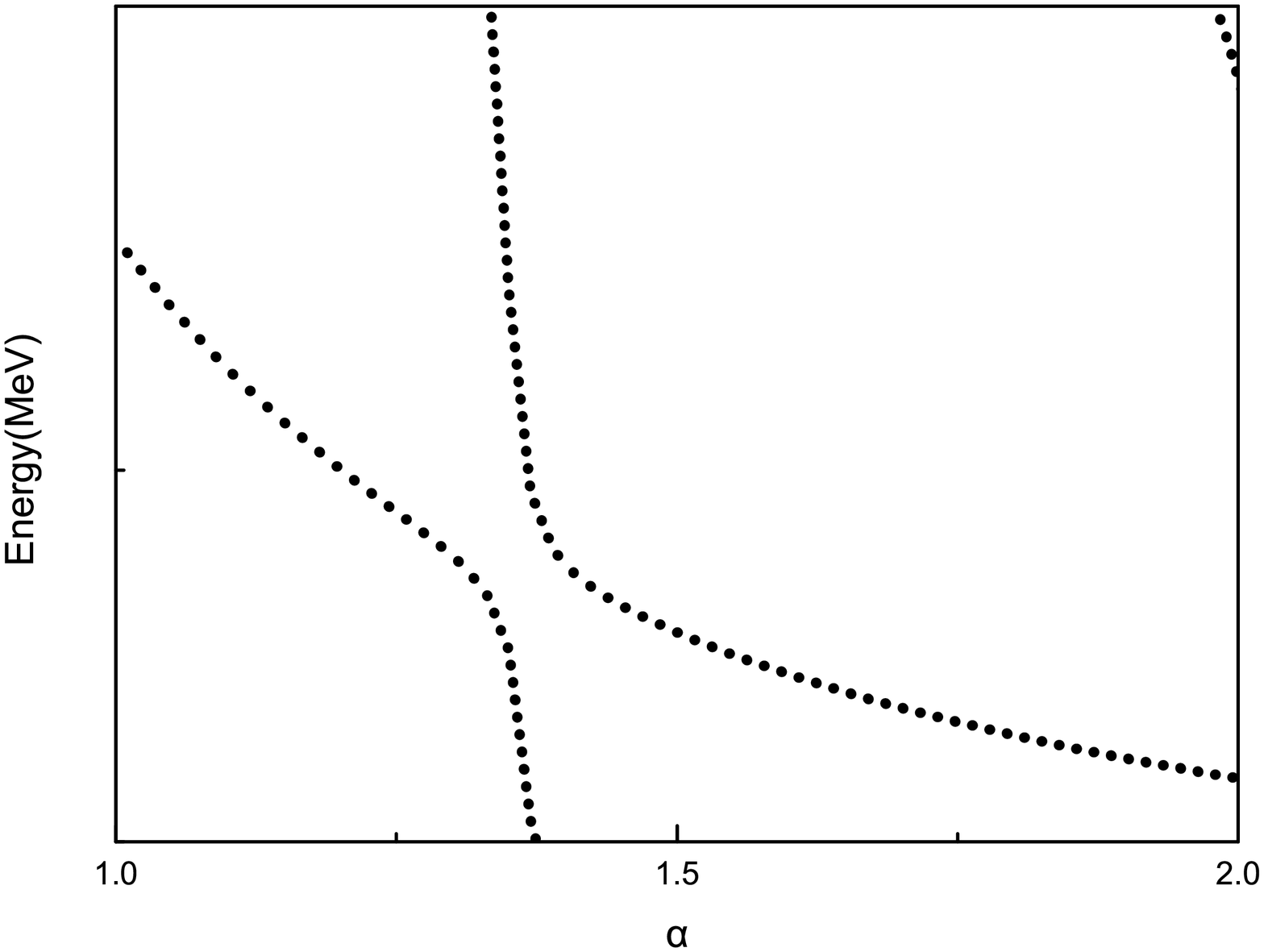}
  \caption{The shape of the resonance in real-scaling method}
  \label{1}
\end{figure}

In Tables III , IV and V, the important calculation results are shown. In each table, columns 2 to 5 represent flavor, spin and color each wave functions
in each channel and the corresponding physical channel of five-quark system. In column 6, the eigen-energy of the each channel is listed and the
theoretical threshold (the sum of the theoretical masses of corresponding baryon and meson) is given in column 7. Column 8 gives the binding energies,
which are the difference between the eigen-energy and the theoretical threshold. Finally, the experimental thresholds (the sum of the experimental
masses of the corresponding baryon and meson) along with corrected energies (the sum of experimental threshold and the binding energy,
$E^{\prime}=E_B+E_{th}^{exp}$) are given in last two columns. With this correction, the calculation error caused by the model parameters in five-quark
calculation can be reduced partly.
\begin{table*}[htb]
\caption{The results for $IJ^P=0\frac{1}{2}^-$. cc1: mixing of color singlet-singlet channels, cc2: mixing of all channels. (unit: MeV)}
\begin{tabular}{cccccccccc}
\hline \hline
~Index~&~~$\psi^{f_i}$~~&~~$\psi^{\sigma_j}$~~&~~$\psi^{c_k}$~~&Physical channel&~~ E ~~& ~~$E^{theo}_{th}$~~ &~~ $E_B$ ~~& ~~$E^{exp}_{th}$ ~~&~~ $E^{\prime}$ ~~\\
\hline
1& $i=1$ & $j=2$ & $k=1$ & $\Lambda\eta_c$ & 4098 & 4098  & 0 & 4100 & 4100 \\
2& $i=1$ & $j=1,2$ & $k=1,2,3$ &  & 4098 & \\
3& $i=1$ & $j=4$ & $k=1$ & $\Lambda J/\psi$ & 4296 & 4296 & 0 & 4213 & 4213  \\
4& $i=1$ & $j=3,4$ & $k=1,2,3$ & & 4296 &  &  &  &  \\
5& $i=2$ & $j=2$ & $k=1$ & $\Lambda_cD_s$ & 4148 & 4148 & 0 & 4250 & 4250  \\
6& $i=2$ & $j=1,2$ & $k=1,2,3$ & & 4148 &  &  &  &  \\
7& $i=2$ & $j=4$ & $k=1$ & $\Lambda_cD_s^*$ & 4340 & 4340 & 0 & 4398 & 4398  \\
8& $i=2$ & $j=3,4$ & $k=1,2,3$ & & 4340 \\
9& $i=3$ & $j=2$ & $k=1$ & $\Xi_cD$ & 4313 & 4318 & -5 & 4335 & 4330  \\
10& $i=3$ & $j=1,2$ & $k=1,2,3$ & & 4312& &-6& &4329  \\
11& $i=3$ & $j=4$ & $k=1$ & $\Xi_cD^*$ & 4480 & 4483  & -3 & 4478 & 4475 \\
12& $i=3$ & $j=3,4$ & $k=1,2,3$ &  & 4478 & & -5 & &4473 \\
13& $i=4$ & $j=1$ & $k=1$ & $\Xi_c'D$ & 4436 & 4439 & -3 & 4443 & 4440  \\
14& $i=4$ & $j=1,2$ & $k=1,2,3$ &  & 4435 & & -4 & &4439  \\
15& $i=4$ & $j=3$ & $k=1$ & $\Xi_c'D^*$ & 4604 & 4604 & 0 & 4586 & 4586  \\
16& $i=4$ & $j=3,4$ & $k=1,2,3$  &  & 4604 & &  & & \\
17& $i=4$ & $j=5$ & $k=1$ & $\Xi_cD^*$ & 4676 & 4676 & 0 & 4652 & 4652  \\
18& $i=4$ & $j=5$ & $k=1,2,3$ & & 4676 \\
cc1 & & & & & 4098 &  & 0  \\
cc2 & & & & & 4098  \\
\hline \hline
\end{tabular}
\end{table*}
\begin{table*}[htb]
\caption{The results for $IJ^P=0\frac{3}{2}^-$. cc1: mixing of color singlet-singlet channels, cc2: mixing of all channels. (unit: MeV)}
\begin{tabular}{cccccccccc}
\hline \hline
~Index~&~~$\psi^{f_i}$~~&~~$\psi^{\sigma_j}$~~&~~$\psi^{c_k}$~~&Physical channel&~~ E ~~& ~~$E^{Theo}_{th}$~~ &~~ $E_B$ ~~& ~~$E^{Exp}_{th}$ ~~&~~ $E^{\prime}$ ~~\\
\hline
1& $i=1$ & $j=7$ & $k=1$ & $\Lambda J/\psi$ & 4296 & 4296  & 0 & 4213 & 4213 \\
2& $i=1$ & $j=6,7$ & $k=1,2,3$ &  & 4296   \\
3& $i=2$ & $j=7$ & $k=1$ & $\Lambda_c D_s^*$ & 4340 & 4340 & 0 & 4398 & 4398  \\
4& $i=2$ & $j=6,7$ & $k=1,2,3$ & & 4340 &  &  &  &  \\
5& $i=3$ & $j=7$ & $k=1$ & $\Xi_c D^*$ & 4481 & 4483 & -2 & 4478 & 4476  \\
6& $i=3$ & $j=6,7$ & $k=1,2,3$ & & 4480 & & -3 & & 4475  \\
7& $i=4$ & $j=7$ & $k=1$ & $\Xi_c'D^*$ & 4600 & 4604 & -4 & 4586 & 4582  \\
8& $i=4$ & $j=6,7$ & $k=1,2,3$ & & 4599 & & -5 & & 4581 \\
9& $i=4$ & $j=8$ & $k=1$ & $\Xi_c^*D$ & 4510 & 4511 & -1 & 4509 & 4508  \\
10& $i=4$ & $j=8$ & $k=1,2,3$ & & 4510 & & -1 & & 4508  \\
11& $i=4$ & $j=9$ & $k=1$ & $\Xi_c^*D^*$ & 4676 & 4676 & 0 & 4652 & 4652  \\
12& $i=4$ & $j=9$ & $k=1,2,3$ & & 4676 & &  & &  \\
cc1 & & & & & 4296 &  & 0  \\
cc2 & & & & & 4296  \\
\hline \hline
\end{tabular}
\end{table*}

\begin{table*}[htb]
\caption{The channels with $IJ^P=0\frac{5}{2}^-$.}
\begin{tabular}{cccccccccc}
\hline \hline
~Index~&~~$\psi^{f_i}$~~&~~$\psi^{\sigma_j}$~~&~~$\psi^{c_k}$~~&Physical channel&~~ E ~~& ~~$E^{Theo}_{th}$~~ &~~ $E_B$ ~~& ~~$E^{Exp}_{th}$ ~~&~~ $E^{\prime}$ ~~\\
\hline
1& $i=4$ & $j=10$ & $k=1$ & $\Xi_c^*D^*$ & 4672 & 4676  & -4 & 4652 & 4648 \\
2& $i=4$ & $j=10$ & $k=1,2,3$ & & 4671 & & -5 & & 4647  \\
\hline \hline
\end{tabular}
\end{table*}
Since we hardly found any bound state in the quantum number of $I=1$ systems, we do not present the results of these systems but focused our analysis
on systems with $IJ^P=0\frac{1}{2}^-$, $IJ^P=0\frac{3}{2}^-$ and $IJ^P=0\frac{5}{2}^-$. The results are analyzed in the following:

(a) For $IJ^P=0\frac{1}{2}^-$ system in Table III: First, the single channel calculations show that there exist weakly bound states
in the $\Xi_cD$, $\Xi_cD^*$ and $\Xi_c'D$ channels. After coupling to respective hidden-color channels, the attractions all increase by a few MeVs,
which is the typical range of binding energy of hadronic molecules. The coupling of all color singlet-singlet channels does not push the lowest
energy of $\Lambda\eta_c$ below its threshold. Then, a full-channel coupling is made and the results show no bound state can be formed.
However, resonances are possible because the attractions exist in the $\Xi_cD$, $\Xi_cD^*$ and $\Xi_c'D$ channels. Thus, real-scaling method is
required to identify resonances.

(b) For $IJ^P=0\frac{3}{2}^-$ system in Table IV: The single channel calculation results show there are weakly bound states in three single channels,
$\Xi_cD^*$, $\Xi_c'D^*$ and $\Xi_c^*D$. Moreover, their respective hidden-color channels increase their attraction a little.
No bound state can be obtained in the color singlet-singlet and full channel coupling calculations. Resonances are still possible.

(c) For $IJ^P=0\frac{5}{2}^-$ system in Table V, there is only one color singlet-singlet channel ($\Xi_c^*D^*$) and there exist a bound state with
binding energy $\sim $5 MeV. Generally attractions exist between two vector hadrons, and the energy of the system is higher than the sum of two
pseudoscalar mesons, resonances are always expected here because of the high angular momentum of the state, which make the coupling between
vector-vector mesons channel and pseudoscalar-pseudoscalar mesons channel via the tensor interaction. Thus, $\Xi_c^*D^*$ can be a good candidate
for pentaquark and more experimental data are needed in the future.

\begin{figure}[h]
  \centering
  \includegraphics[width=9cm,height=7cm]{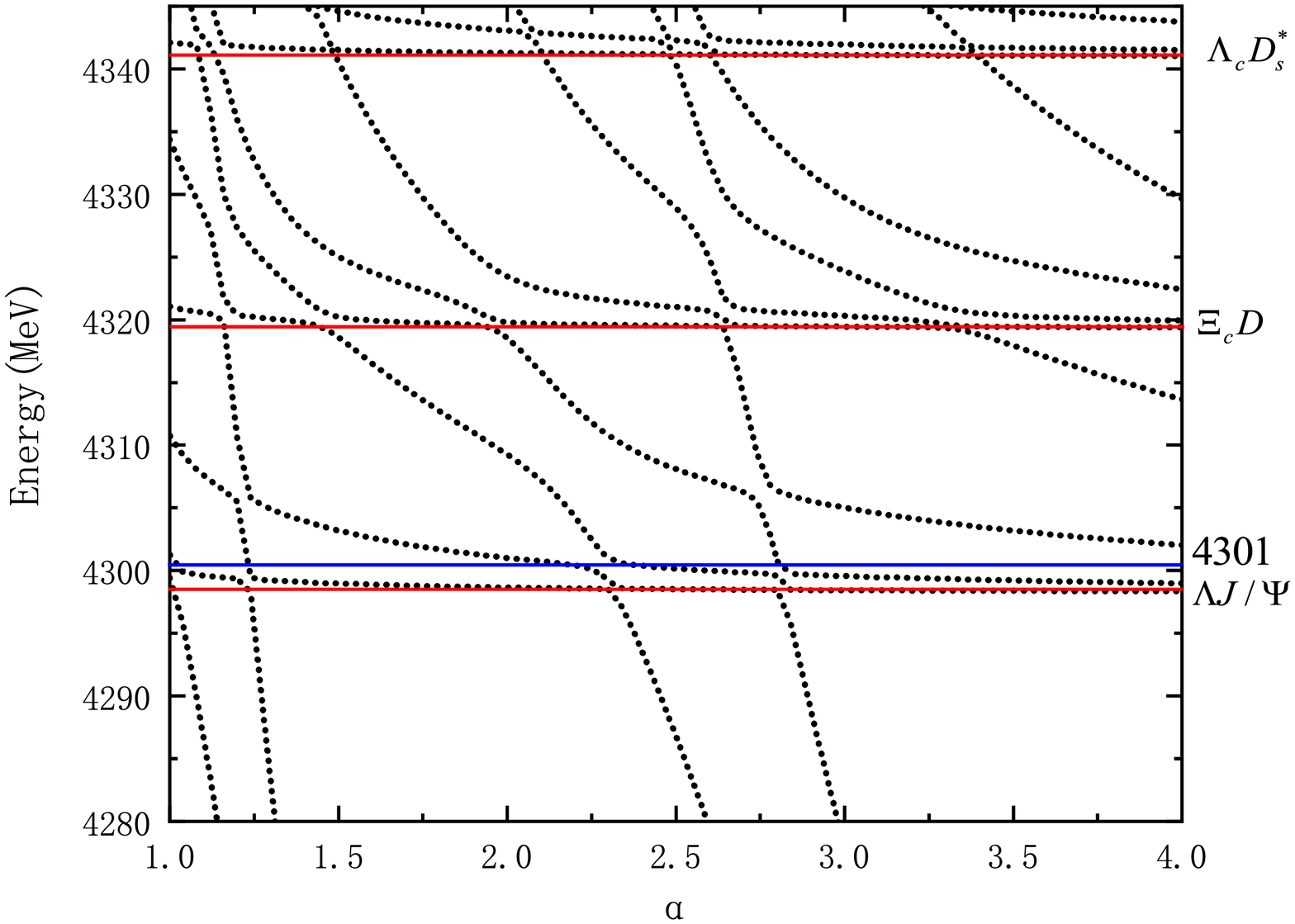}
  \includegraphics[width=9cm,height=7cm]{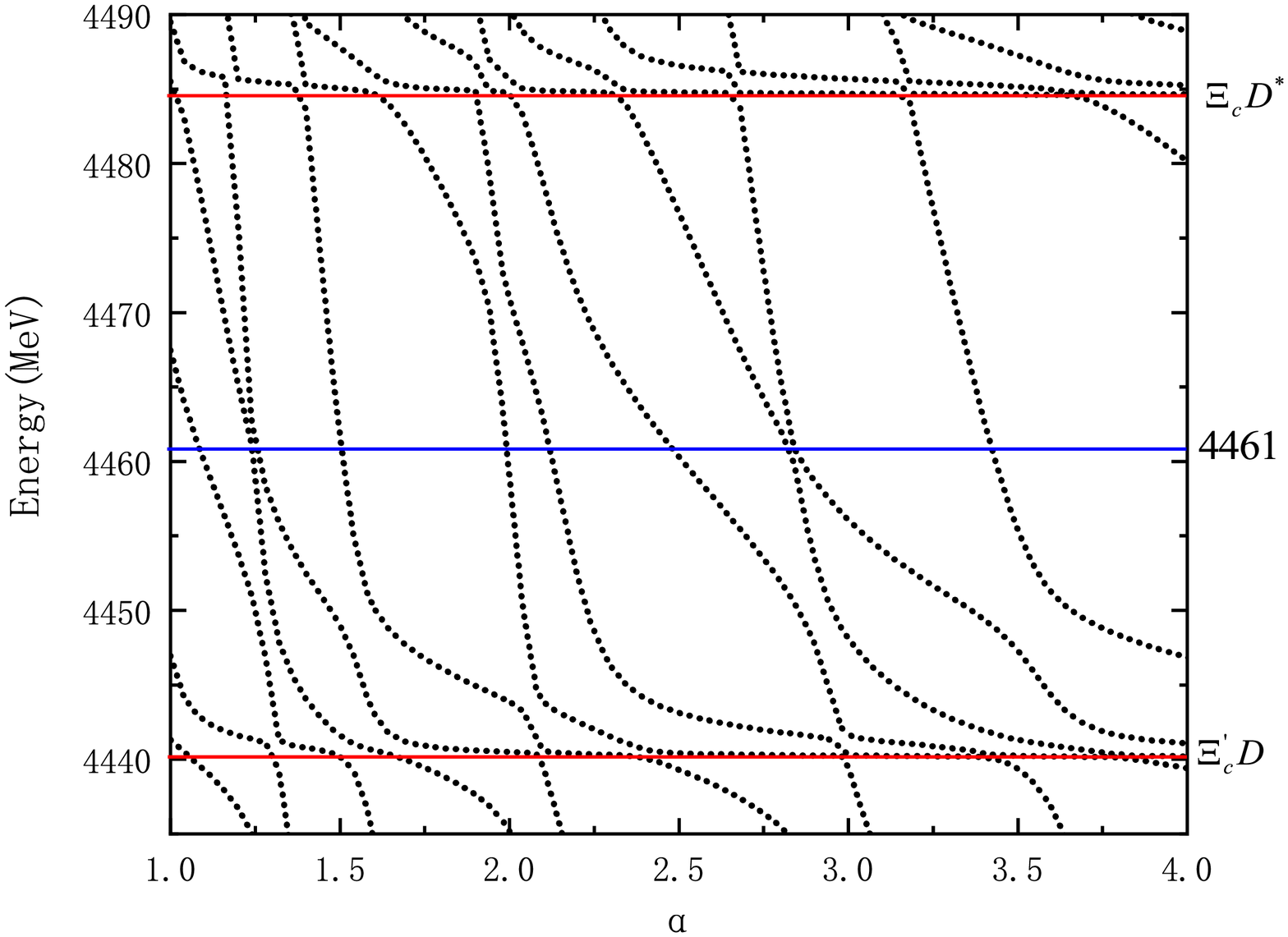}
  \caption{Energy spectrum of $0\frac{1}{2}^-$ system.}
  \label{1}
\end{figure}
\begin{figure}[h]
  \centering
  \includegraphics[width=9cm,height=7cm]{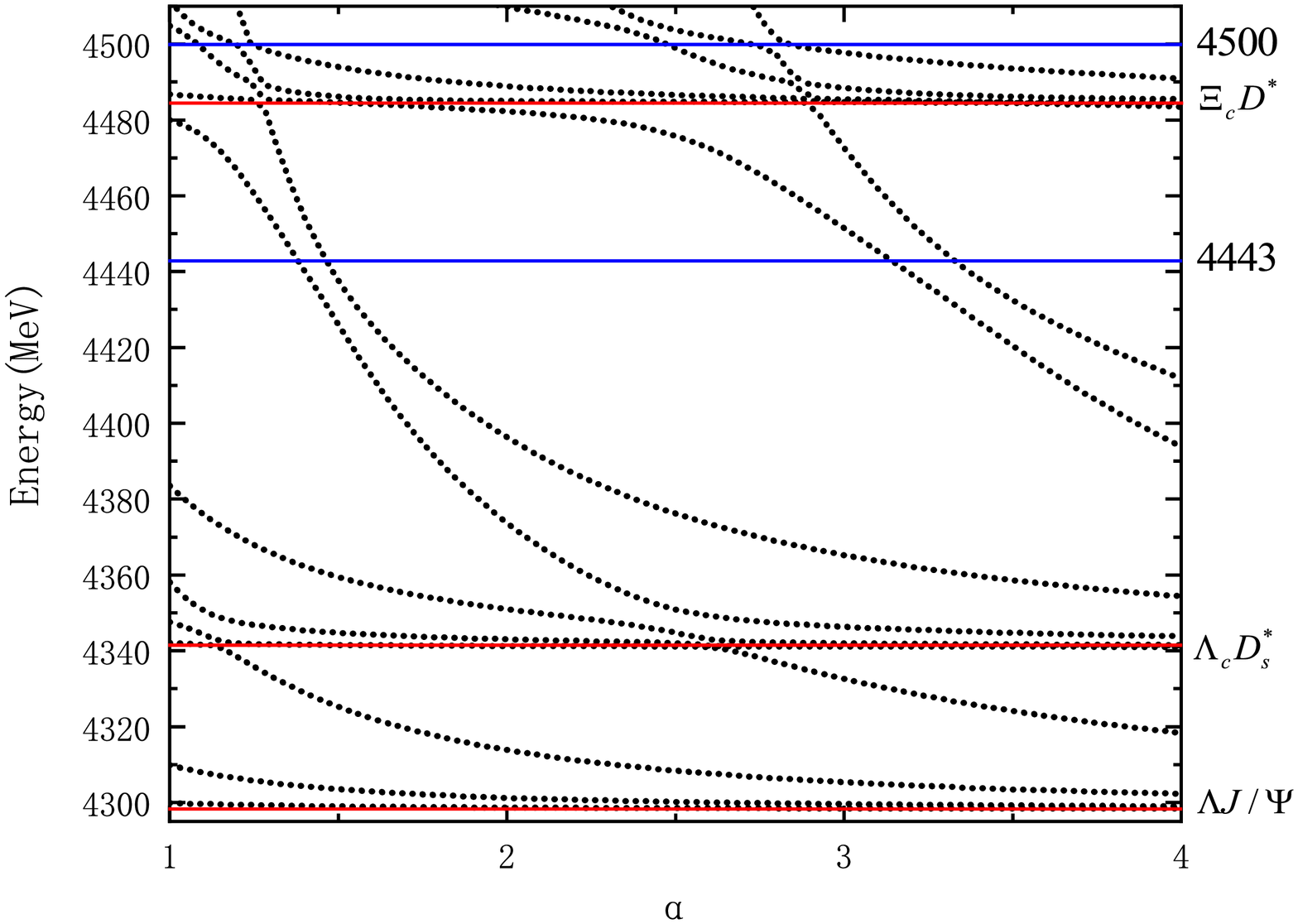}
  \includegraphics[width=9cm,height=7cm]{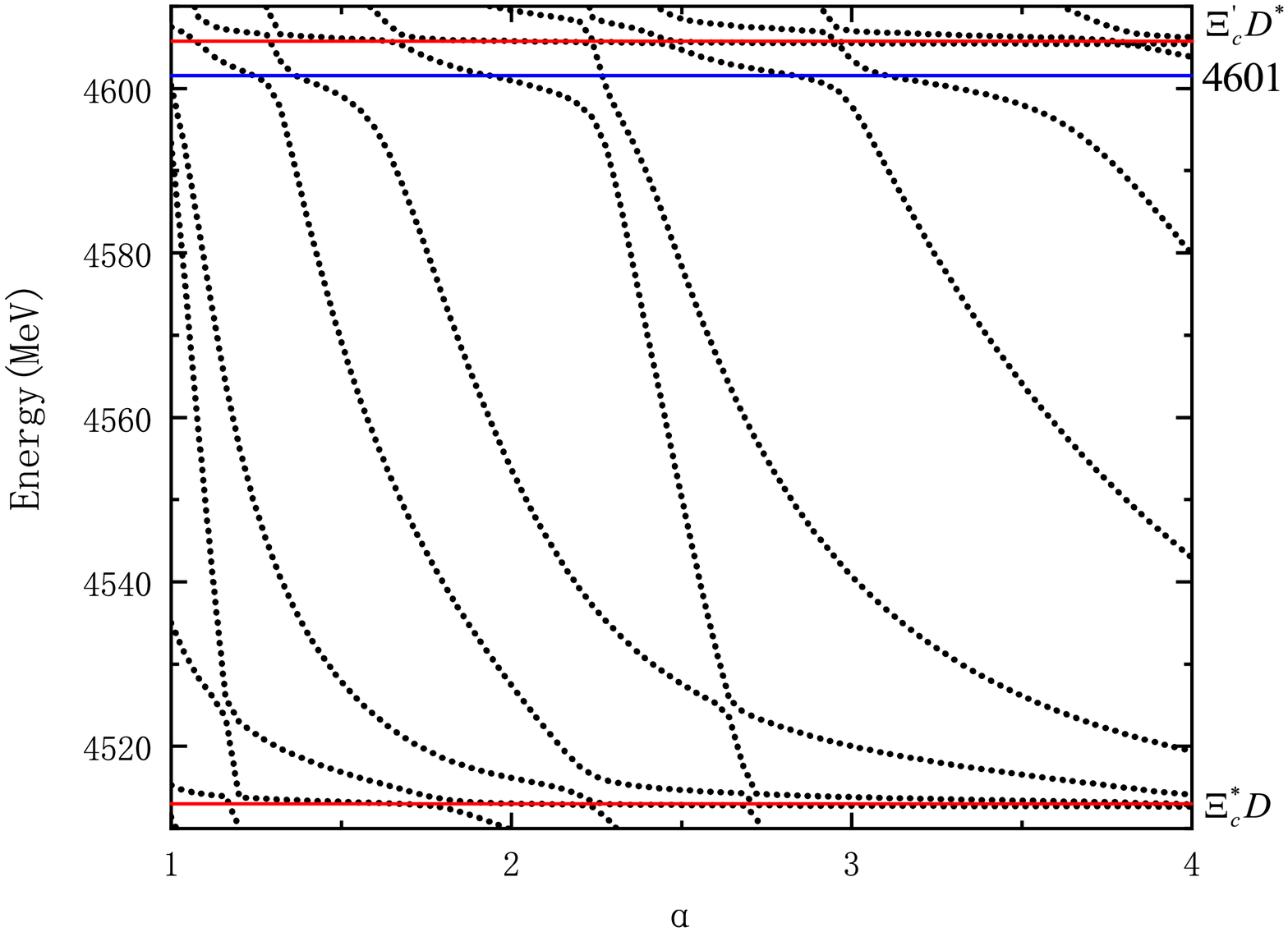}
  \caption{Energy spectrum of $0\frac{3}{2}^-$ system.}
  \label{1}
\end{figure}

To check whether the resonances with quantum numbers $IJ^P=0\frac{1}{2}^-,0\frac{3}{2}^-$ can survive after coupling to the open channels,
the real-scaling method is employed. The results are shown in Figs.~2, 3 and 4. In these figures, the thresholds of all physical
channels appear as horizontal lines and are marked with lines (red lines), tagged with their contents. And for genuine resonances, which appear as
avoid-crossing structure and are marked with blue lines. The continuum states fall off towards their respective threshold states (red horizontal lines).

For $IJ^P=0\frac{1}{2}^-$ system, we get two resonances whose energy are 4301 MeV (main component is $\Xi_cD$) and 4461 MeV (main component is $\Xi_c D^*$). Especially for $P_{cs}(4461)$, which is very close to 4459 MeV, it is a good candidate for $P_{cs}(4459)$ reported by LHCb Collaboration.
In $IJ^P=0\frac{3}{2}^-$ system, there are three resonances, $P_{cs}(4443)$ (main component is $\Xi_cD^*$), $P_{cs}(4500)$
(main component is $\Xi_c^*D$) and $P_{cs}(4601)$ (main component is $\Xi_c'D^*$). Finally, in $IJ^P=0\frac{5}{2}^-$ system, there is only one bound
state marked horizontally under its threshold ($P_{cs}(4671)$), it will turn to a narrow resonance state after coupling to $\Xi_cD$ via tensor interaction.

It is worth mentioning that in case we use real-scaling method to identify resonances, we will also calculate the composition of the possible resonances
to find the mechanism of the formation the resonances. The main component of a genuine resonance should be bound state channels in the single channel
or coupling channel (the main component channel and other channels with energies higher than the main component channel) calculations.
In Figs.~2, 3, in addition to the resonances found, there are other avoid-crossing structures such as 4306 MeV in $IJ^P=0\frac{1}{2}^-$ system
and 4525 MeV in $IJ^P=0\frac{3}{2}^-$ system. However, their components are open channels, which means the avoid-crossing structures are formed due to
the difference in the decay slope of different open channels.

For resonances, the partial widths of two mesons strong decay can be extracted from theses figures.
The decay width of resonance to possible open channels of two mesons is obtained by the following formula
\begin{eqnarray}
&&\Gamma=4V(\alpha)\frac{\sqrt{(k_r \times k_c)}}{|k_r-k_c|},
\end{eqnarray}
where $V(\alpha)$ is the minimum energy difference, while $k_c$ and $k_r$ stand for the slopes of scattering state and resonance state respectively.
More details can be found in Ref.~\cite{35}.

\begin{figure}[t]
  \centering
  \includegraphics[width=9cm,height=7cm]{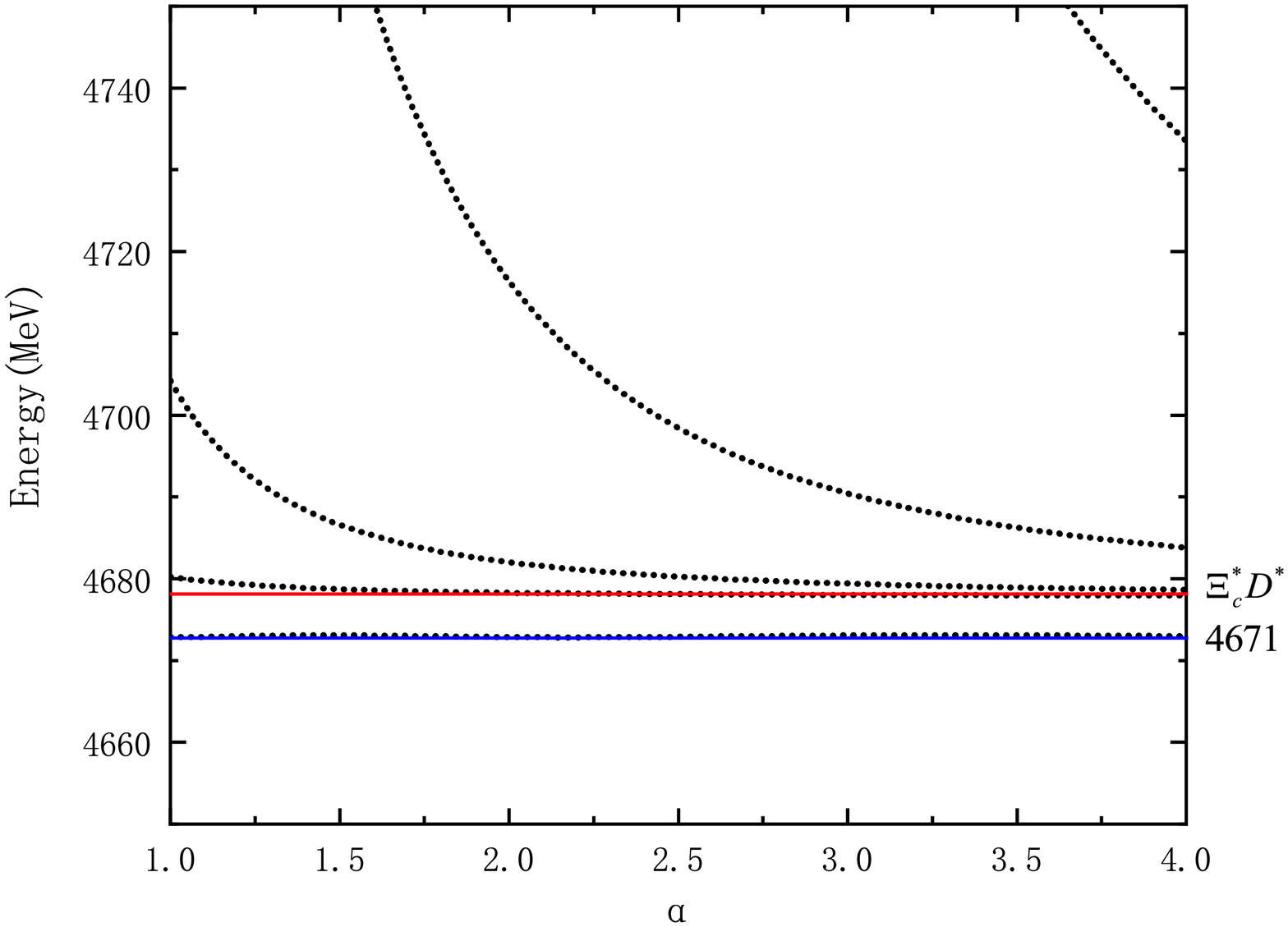}
  \caption{Energy spectrum of $0\frac{5}{2}^-$ system.}
  \label{1}
\end{figure}

To extract the partial decay to specific channel, more calculations are needed. For example, to obtain the partial decay width of resonance $P_{cs}(4461)$
in $IJ^P=0\frac{1}{2}^-$ system, the channel coupling calculations with three bound state channels, $\Xi_cD$, $\Xi_cD^*$ and $\Xi_c'D$ and one of four open
channels: $\Lambda\eta_c$, $\Lambda J/\psi$, $\Lambda_cD_s$, and $\Lambda_cD_s^*$ are performed, the results are shown in Fig. 5. From Fig. 5(a), one can
extract the partial decay width of $P_{cs}(4461)$ to $\Lambda\eta_c$ and so on. In different channel coupling calculations, the obtained resonance energies
are slightly different, from 4453 MeV to 4458 MeV. It is due to the finite model space used, the avoid-crossing structures appear at different scaling factors.
Finally, all possible resonances and their respective partial decay widths
and total decay width to two ground hardrons are shown in the Table VI and VII. The main component of each resonance and corrected energy are also given
in the tables.

\begin{figure}[h]
  \centering
  \includegraphics[width=9.5cm]{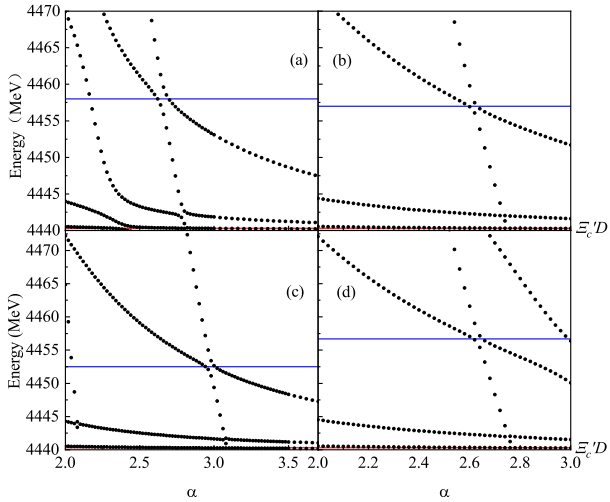}
  \caption{Energy spectrum of $P_{cs}(4461)$ in four channels coupling, three bound state channels: $\Xi_cD$, $\Xi_cD^*$ and $\Xi_c'D$,
  and one open channel for $0\frac{1}{2}^-$ system. The open channel is (a) $\Lambda\eta_c$, (b) $\Lambda J/\psi$, (c) $\Lambda_cD_s$, and (d) $\Lambda_cD_s^*$.}
  \label{1}
\end{figure}

\begin{table}[h]
\caption{The decay width of $P_{cs}$ states in $0\frac{1}{2}^-$ system (unit: MeV).}
\begin{tabular}{ccccccc}
\hline
~~~Mode~~~~ & ~~$P_{cs}(4301)$~~ & ~~$P_{cs}(4461)$~~        \\ \hline
~Total Width~ & 4.0  &8.9  &  \\\hline
$\Lambda\eta_c$ & 2.5   & 3.7 & \\\hline
$\Lambda J/\psi$ & 1.5 & 1.2  &    \\\hline
$\Lambda_cD_s$ & ? & 1.5 &  \\\hline
$\Lambda_cD_s^*$  & ---   & 2.5 &    \\\hline
Main Comp. &  $\Xi_cD$ & $\Xi_c D^*$ &    \\\hline
$E^{\prime}$ &  4318  & 4458  &    \\\hline
\label{width1}
\end{tabular}

~

\caption{The decay width of $P_{cs}$ states in $0\frac{3}{2}^-$ system (unit: MeV).}
\begin{tabular}{ccccccc}
\hline \hline
Mode &~$P_{cs}(4443)$~ & $P_{cs}(4500)$    & $P_{cs}(4601)$    \\ \hline
~Total Width~ &  24.4 &  7.3&  5.9& \\\hline
$\Lambda J/\psi$ & 19.6& 1.7 &  3.3&   \\\hline
$\Lambda_cD_s^*$ &  4.8 & 5.6 & 2.6 &   \\\hline
Main Comp. &  $\Xi_cD^*$ & $\Xi_c^*D$ & $\Xi_c^{\prime}D^*$ &   \\\hline
$E^{\prime}$ & ~4444~ &~4497~ &~4585~ &   \\\hline
\hline \label{width2}
\end{tabular}
\end{table}
From Tables \ref{width1}-\ref{width2}, we can see that, in $0\frac{1}{2}^-$ system,  the total decay width of $P_{cs}(4301)$ is 4.0 MeV and
we found its decay width to channel $\Lambda\eta_c$ is very narrow. However, we found no typical avoid-crossing structures when we calculated its decay width to channel $\Lambda_cD_s$, which is a problem that needs further study. For $P_{cs}(4461)$, the decay width is 8.9 MeV and $\Lambda\eta_c$ is the main decay
channel. In $0\frac{3}{2}^-$ system, the decay width of two resonances $P_{cs}(4500)$ and $P_{cs}(4601)$ are 7.3 MeV and 5.9 MeV, the main decay
channel of $P_{cs}(4500)$ is $\Lambda_cD^*_s$, and $P_{cs}(4601)$ has comparable decay widths to $\Lambda_cD^*_s$ and $\Lambda J/\psi$.
For $P_{cs}(4443)$, the decay width is 24.4 MeV and its main decay channel is $\Lambda J/\psi$. For $0\frac{5}{2}^-$ system, there is only resonance $P_{cs}(4671)$, which is 5 MeV
under its threshold and the decay width is  estimated to be 5-15 MeV due to decay widths of its constituents, $\Xi_c^*$ ($\Gamma_{\Xi_c^*\longrightarrow\Xi_c\pi}$$\sim$2.35 MeV) and $D^*$ ($\Gamma_{D^*\longrightarrow D\pi}$$\sim$2 MeV) and the decay width to
$\Xi_cD$, $1\sim 10$ MeV, via tensor interaction. Considering the deviation of quark model calculations and the uncertainty of experimental
results, $P_{cs}(4461)$ in $0\frac{1}{2}^-$ system can be identified as  $P_{cs}(4459)$ reported by LHCb Collaboration.

\section{Summary}
In this paper, the hidden-charm pentaquark systems with single strangeness is investigated in chiral quark model. The calculation shows that there are
several states in systems $0\frac{1}{2}^-$, $0\frac{3}{2}^-$ and $0\frac{5}{2}^-$, which means good resonances can be formed. Real-scaling method are used
to check the genuine resonances and study their decay width. In $0\frac{1}{2}^-$ system, two resonances $P_{cs}(4301)$ and $P_{cs}(4461)$ have been found. In particular, $P_{cs}(4461)$, whose main component is $\Xi_cD^*$, is regarded as a good candidate for $P_{cs}(4459)$ recently reported
by LHCb Collaboration. Other possible pentaquarks are also predicted. Three states, $P_{cs}(4443)$ ,$P_{cs}(4500)$ and $P_{cs}(4601)$ are found in system $0\frac{3}{2}^-$.
It is worth mentioning that there are $P_{cs}(4461)$ and $P_{cs}(4443)$ in our calculations, which can form a two-peak structure composed of $\Xi_cD^*$ states
with quantum number of $0\frac{1}{2}^-$ and $0\frac{3}{2}^-$, similar to $P_{c}(4440)$ and $P_{c}(4457)$. Finally, only one molecule state $\Xi_c^*D^*$
exists in system $0\frac{5}{2}^-$ and it is a good candidate for heavy pentaquark with high spin with its decay width is in range of 5 MeV to 15 MeV.

From the results, one can see that the behaviors of $P_{cs}$ systems are similar to that of $P_c$ systems. Due to the success of chiral quark model on
$P_{c}$ states~\cite{28}, it is expected that the possible resonances proposed above can be searched in future experiments.

~

\section*{Acknowledgments}
The work is supported partly by the National Natural Science Foundation of China under Grant
Nos. 11775118, and 11535005.

\end{document}